\documentclass{article} 
\usepackage{iclr2020_conference,times}

\usepackage{amsmath,amsfonts,bm}

\def\eqref#1{equation~\ref{#1}}

\def\1{\bm{1}}

\DeclareMathAlphabet{\mathsfit}{\encodingdefault}{\sfdefault}{m}{sl}
\SetMathAlphabet{\mathsfit}{bold}{\encodingdefault}{\sfdefault}{bx}{n}

\usepackage{amsmath}
\usepackage{amsthm}
\usepackage{amssymb}
\usepackage{graphicx, subcaption}

\usepackage{hyperref}
\usepackage{url}

\title{Bringing PDEs to JAX with forward and reverse modes automatic differentiation}

\author{Ivan Yashchuk\\
Department of Computer Science, Aalto University\\
VTT Technical Research Centre of Finland\\
\texttt{ivan.yashchuk@\{aalto,vtt\}.fi} \\
}

\newcommand\rurl[1]{%
  \href{http://#1}{\nolinkurl{#1}}%
}

\iclrfinalcopy 
\begin{document}

\maketitle

\begin{abstract}
Partial differential equations (PDEs) are used to describe a variety of physical phenomena.
Often these equations do not have analytical solutions and numerical approximations are used instead. One of the common methods to solve PDEs is the finite element method.
Computing derivative information of the solution with respect to the input parameters is important in many tasks in scientific computing.
We extend JAX automatic differentiation library with an interface to Firedrake finite element library.
High-level symbolic representation of PDEs allows bypassing differentiating through low-level possibly many iterations of the underlying nonlinear solvers.
Differentiating through Firedrake solvers is done using tangent-linear and adjoint equations.
This enables the efficient composition of finite element solvers with arbitrary differentiable programs.
The code is available at
\rurl{github.com/IvanYashchuk/jax-firedrake}.
\end{abstract}

\section{Introduction}


There is a growing interest in differentiable physical simulators and including them as components in machine learning systems.
Recent work in this field includes either implementing physical simulators from scratch in popular deep learning frameworks \citep{phiflow} or developing new software specifically for physical simulations \citep{difftaichi}.
In the case of simulators consisting of the partial differential equation (PDE) solving
differentiating through all the operations of the PDE solver is not scalable and heavy memory-bound.
For PDEs, explicit expressions are available for deriving efficient functions needed for automatic differentiation (AD).
In this work, we integrate Firedrake PDE library with JAX AD library.
As a result, it is possible to compose Firedrake models with arbitrary programs that JAX can differentiate.


\section{Background}
The variational formulation also known as weak formulation allows finding the solution to problems modeled with PDEs using an integral form. Having the integral form gives
simpler equations to solve using linear algebra methods over a vector space of infinite
dimension or functional space.
In finite element methods (FEM), the variational formulation of the problem is transformed into a system of nonlinear equations for unknown finite element function coefficients that can be solved numerically.

Let a system of PDEs that describe the physics of the problem of interest be written as
\begin{equation}\label{abstract-pde}
F(u, m) = 0,
\end{equation}
where $u$ is the solution and $m$ represents parameters that affect the solution. The solution $u$ can be written as an implicit function $u(m)$ of the parameters $m$,
then we can formulate its derivative $\frac{d u}{d m}$.
In this work we consider the non-linear systems of PDEs that are discretized using the finite element method.

Let $J(u, m)$ be a functional of interest, it represents any quantity that depends on $u$ and $m$. Then the problem is computing the derivatives $\frac{d J(u(m), m)}{d m}$. It can be done using finite difference methods, tangent-linear or adjoint approaches.


The tangent-linear system is the same idea as the forward mode of automatic differentiation,
while the adjoint approach corresponds to the reverse mode automatic differentiation.
In \citep{Farrell2013}, authors describe the techniques that
can be used to derive possibly time-dependant tangent-linear and adjoint models.
Having the solution to the tangent-linear system we can evaluate the gradient of any functional. However, in many applications the functional is fixed and the goal is to calculate the derivative with respect to any parameter the chosen functional depends on. In this situation, the better approach is to use the adjoint equations for derivative computation.
Higher-order derivatives can be derived by composing tangent-linear and adjoint models \citep{Maddison2019}.

\paragraph{Firedrake library}
Firedrake is an automated system for the solution of partial differential equations using the finite element method. It uses the UFL language to express variational problems \citep{Rathgeber2016, ufl_paper}. Then this high-level problem representation is compiled into low-level C code for assembling vectors and matrices for the nonlinear solver. Nonlinear and linear solvers are based on PETSc library \citep{petsc-web-page}.
UFL representation of the residual equation makes it straightforward to generate both tangent-linear and adjoint equations using built-in automatic differentiation.
However, pure UFL implementation of derivative calculation is restricted only to one variational problem.
Time-dependent problems are solved as a sequence of variational problems and
to differentiate through this sequence program execution should be traced.
Two libraries interface directly with Firedrake for automated derivative computation:
dolfin-adjoint \citep{Mitusch2019} and tlm\_adjoint \citep{Maddison2019}.
So it is not necessary to use dolfin-adjoint and relying only on UFL for adjoint and tangent-linear derivation is possible if the Firedrake model is a single variational problem or time-stepping is implemented in JAX. However, we chose to depend on dolfin-adjoint library here as it also makes possible to calculate shape derivatives for domain optimization and boundary conditions derivatives.

\section{Automatically differentiating PDE solvers in JAX}
JAX is a numerical computing library that includes forward and reverse mode automatic differentiation.
JAX uses terminology and concepts from differential geometry, namely pushforward map for the forward mode AD and pullback map for the reverse mode AD.
These maps can also be described in terms of the Jacobian: The pushforward is Jacobian-vector product (JVP), and pullback is Jacobian-transpose-vector product, or vector-Jacobian product (VJP).

Given a function $f: \mathbb{R}^n \to \mathbb{R}^m$, the Jacobian matrix of $f$ evaluated at an input point $x \in \mathbb{R}^n$, denoted $\partial f(x)$,
is often thought of as a matrix of partial derivatives of size $m \times n$. Alternatively, $\partial f(x)$ represents a linear map, which maps the tangent space of the domain of $f$ at the point $x$
to the tangent space of the codomain of $f$ at the point $f(x)$: 
\begin{equation*}
\partial f(x) : \mathbb{R}^n \to \mathbb{R}^m.
\end{equation*}
This map is called the pushforward map of $f$ at $x$.
Given input point $x \in \mathbb{R}^n$ and a tangent vector $v \in \mathbb{R}^n$, we get back an output tangent vector in $\mathbb{R}^m$. This mapping, from $(x,v)$ pairs to output tangent vectors, is called the Jacobian-vector product, and written as
\begin{equation*}
(x, v) \mapsto \partial f(x) v.
\end{equation*}

The pullback map of $f$ at $x$ is
\begin{equation*}
\partial f(x)^* : \mathbb{R}^m \to \mathbb{R}^n.
\end{equation*}
The Jacobian-transpose-vector product is then
\begin{equation*}
(x, v) \mapsto \partial f(x)^* v.
\end{equation*}

Referring back to the tangent-linear model in the context of PDEs, there Jacobian-vector product corresponds to
\begin{equation}
(m, v) \mapsto \frac{d u}{d m} v := \dot{u}_v \quad \in \mathbb{R}^m,
\end{equation}
where $\dot{u}_v$ is the solution to the following tangent-linear equation with right hand side defined as
the derivative of $F(u, m)$ with respect to $m$ in the direction $v \in \mathbb{R}^n$:
\begin{equation}\label{eq:tlm_ad}
\frac{\partial F(u,m)}{\partial u} \dot{u}_v = - \frac{\partial F(u,m)}{\partial m} \cdot v.
\end{equation}
Note that compared to the standard tangent-linear equations the unknown
in the \eqref{eq:tlm_ad} is not the full solution Jacobian matrix but its multiplication with a vector,
making the problem easier to solve.

Jacobian-transpose-vector product function is defined as
\begin{equation*}
(m, v) \mapsto  \frac{d u}{d m}^* v := - \frac{\partial F(u, m)}{\partial m}^* \cdot \lambda_v \quad \in \mathbb{R}^n,
\end{equation*}
where $\lambda_v$ is the solution to the following adjoint equation with right hand side given by the cotangent vector $v \in \mathbb{R}^m$:
\begin{equation}\label{eq:adjoint_ad}
\frac{\partial F(u, m)}{\partial u}^{*} \lambda_v = v.
\end{equation}
Compared to the standard adjoint equations here we do not need any information about the functional and it is AD's system responsibility to compose VJPs for reverse pass and supply cotangent vectors.

Internally, JAX functions are called Primitives and for each Primitive Jacobian-vector products (JVP), vector-Jacobian products (VJP), batching and just-in-time compilation rules are implemented.
To make JAX work with external functions we need to implement new Primitives together with functions that define vector-Jacobian and Jacobian-vector products.

We have implemented a new Primitive function for the Firedrake solver that transforms inputs and outputs to appropriate data types and registers associated JVP and VJP functions that solve tangent-linear and adjoint equations respectively.


\section{Numerical examples}

For demonstrating the implementation we consider the Poisson equation as the forward model problem.
Let $\Omega \subset \mathbb{R}^n, n\in \{1,2,3\}$ be an open, bounded domain and consider the following problem:
\begin{equation}\label{eq:poisson-strong}
\begin{split}
\quad -\nabla\cdot(\kappa \nabla u) &= f \text{ in }\Omega,\\
u &= 0 \text{ on }\partial\Omega,
\end{split}
\end{equation}
where $u: \Omega \to \mathbb{R}$ is the unknown temperature,
$\kappa \in \mathbb R$ is the thermal conductivity, $f: \Omega \to \mathbb R$ is the source term ($f(x)>0$  corresponds to heating and $f(x)<0$ corresponds to cooling).

The variational form of the \eqref{eq:poisson-strong}:
Find $u\in H_0^1(\Omega)$ such that
\begin{equation}\label{eq:poisson-weak}
(\kappa \nabla u,\nabla v)_{L^2(\Omega)} - (f,v)_{L^2(\Omega)} = 0, \text{ for all } v\in H_0^1(\Omega),
\end{equation}
where $H_0^1(\Omega)$ is the space of functions vanishing on $\partial\Omega$ with square integrable
derivatives. $(\cdot\,,\cdot)_{L^2(\Omega)}$ denotes the $L^2$-inner product.

\paragraph{Optimal control of the Poisson equation}
We solve the standard problem in PDE-constrained optimization: the optimal control of the Poisson equation. Physically, this problem can be interpreted as finding the best heating or cooling of a surface to achieve a desired temperature profile.
The problem is to minimize the following functional
\begin{equation}
\min_{f} J(f):=\frac{1}{2}\int_\Omega (u - u_d)^2\, dx + \frac{\gamma}{2} \int_\Omega f^2\,ds,
\end{equation}
where $u$ is the solution to the Poisson \eqref{eq:poisson-weak}, $u_d$ is the desired temperature profile, $f$ is the unknown control function, $\gamma$ is the regularization parameter.
Additionally, $f$ is constrained to bounds $a, b$ such that $a \le f \le b$.

The unknown function $f$ is discretized in a finite element space such that values of $f$ at each cell of the mesh are the optimization parameters. As the mesh is refined the number of parameters in the optimization problem increases.

For our example we take the desired temperature to be $u_d= \frac{1}{2\pi^2}\frac{1}{1+4\gamma \pi^4}\sin(\pi x) \sin(\pi y)$, $\gamma=10^{-6}$, and $f$ is bounded between $0$ and $0.8$.
L-BFGS-B optimizer from SciPy library \citep{2020SciPy-NMeth} is used with the gradient values calculated by JAX.
Convergence is achieved after 38 iterations with gradient norm tolerance $10^{-10}$.

\paragraph{Coefficient field inversion with neural network representation}
Representing the model parameters at each
point in space quickly leads to a large number of model
parameters.
The neural
network can be used as an approximation to the spatially varying coefficients characterized by the weights of the neural network. The optimization problem is then posed in the space of network-parameters rather than at each cell of the computational grid.
In the task of topology optimization, it was shown that neural network representation of the solution to the optimization problem helps to find a better design in many cases \citep{1909.04240}.
In \citep{1712.09685}, the authors studied neural network representation stability for inverse problems.

In this example, we demonstrate the use of neural nets to parameterize inputs of Firedrake solver and demonstrate the differentiability of the pipeline.
The problem is to minimize the following functional
\begin{equation}
\min_{\kappa} J(\kappa):=\frac{1}{2}\int_\Omega (u-u_m)^2\, dx + \frac{\gamma}{2}\int_\Omega|\nabla \kappa|^2\,dx,
\end{equation}
where $u$ is the solution to the Poisson \eqref{eq:poisson-weak}, $u_m$ is the noisy temperature measurement, $\kappa$ is the unknown material coefficient field, $\gamma$ is the regularization parameter.

\begin{figure}[b]
\centering
\begin{subfigure}{.3\textwidth}
\centering
\includegraphics[width=\linewidth]{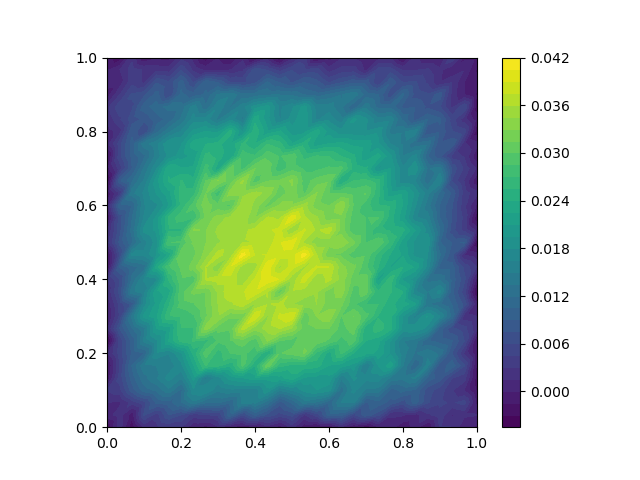}
\caption*{Noisy temperature field}
\end{subfigure}%
\begin{subfigure}{.3\textwidth}
\centering
\includegraphics[width=\linewidth]{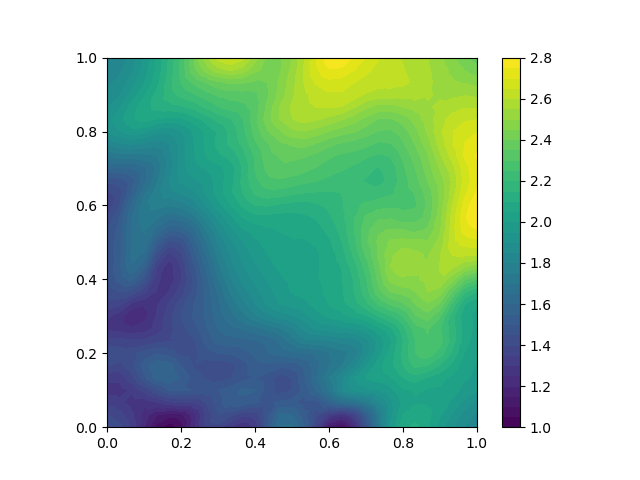}
\caption*{Inverted $\kappa$, FEM}
\end{subfigure}%
\begin{subfigure}{.3\textwidth}
\centering
\includegraphics[width=\linewidth]{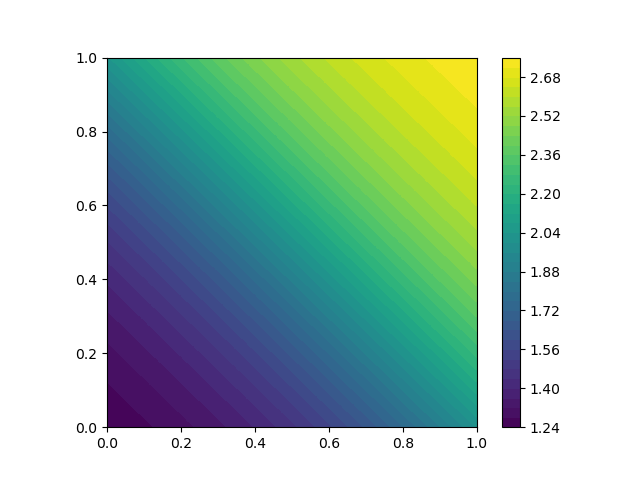}
\caption*{Inverted $\kappa$, NN}
\end{subfigure}%
\caption{Conductivity field inversion from noisy measurements using FEM and NN representations of optimization parameters.}
\label{fig:inverse-poisson}
\end{figure}

Here, for the simplicity, we choose feed-forward neural network with single hidden layer with 10 neurons and tanh activation function. We set up synthetic measurements temperature data with spatially varying coefficient $\kappa = 1 + x + y$ and adding small noise to the true temperature state.
\begin{table}[t]
\centering
\begin{tabular}{|l|l|l|l|l|}
\hline
    & \# parameters & \# iterations & $\|\kappa_{opt}-\kappa_{true}\|$ & $\|u_{opt}-u_{true}\|$ \\ \hline
FEM & 981          & 31           & 1.356112e-1                       & 2.711260e-4              \\ \hline
NN  & 47           & 89           & 5.362389e-2                       & 1.999892e-4              \\ \hline
\end{tabular}
\caption{\label{tab:inverse-res}Performance comparison between FEM and NN representation of the
coefficient.}
\end{table}
L-BFGS-B optimizer is used with gradient norm tolerance $10^{-6}$.
Number of iterations and $L^2$-norm of the difference between the optimal solution and the true one are summarized in Table \ref{tab:inverse-res}.
In this problem, finite element representation overfits to the noise present in the measurements
while neural network representation can recover material coefficient function closer to the true function and visually seems to be unaffected by the noise (Figure \ref{fig:inverse-poisson}).

\section{Conclusions}
We describe an extension to JAX that allows a seamless inclusion of PDE solvers written in Firedrake into arbitrary JAX differentiable programs, including, but not limited to, deep learning models, Bayesian probabilistic models.
Ongoing work targets further integration of Firedrake code with JAX allowing just-in-time compilation and GPU computing and better interoperability of JAX parallelism with Firedrake's MPI parallelism.

\newpage
\bibliography{iclr2020_conference}
\bibliographystyle{iclr2020_conference}

\newpage
\appendix
\section{Derivations of tangent-linear and adjoint equations}


Begin with the non-linear system of equations given by \eqref{abstract-pde}.
Then, taking the total derivative of the above equation with respect to the
parameter $m$ gives
\begin{equation}
\frac{d } {d m} F(u, m) = \frac{\partial F(u,m)}{\partial u} \frac{d u}{d m} + \frac{\partial F(u,m)}{\partial m}\frac{d m}{d m} = 0
\end{equation}
and re-arranging gives the tangent-linear equation associated with the PDE \eqref{abstract-pde}
\begin{equation}\label{eq:tlm1}
\frac{\partial F(u,m)}{\partial u} \frac{d u}{d m} = - \frac{\partial F(u,m)}{\partial m}.
\end{equation}
The tangent equation is always linear.


Consider a functional $J(u, m)$. Let $\widehat{J}(m):= J(u(m), m)$ be a pure function of $m$.
Applying the chain rules gives the expression for the gradient
\begin{equation}\label{eq:gradJ1}
\frac{d }{d m} \widehat{J}(m) = \frac{\partial J(u, m)}{\partial u} \frac{d u}{d m} + \frac{\partial J(u, m)}{\partial m}.
\end{equation}
Having the solution to the tangent-linear \eqref{eq:tlm1} we can evaluate the gradient of any functional $\widehat{J}$. However, in many applications the functional is fixed and the goal is to calculate the derivative with respect to any parameter the chosen functional depends on. In this case, the alternative is the adjoint approach.

Suppose the tangent-linear system is invertible. Then rewrite the solution to the \eqref{eq:tlm1} as
\begin{equation}
\frac{d u}{d m} = - \left(\frac{\partial F(u,m)}{\partial u}\right)^{-1} \frac{\partial F(u,m)}{\partial m}
\end{equation}
and substitute $\frac{d u}{d m}$ into the expression for the gradient of $\widehat{J}$:
\begin{equation}
\frac{d \widehat{J}(m) }{d m} = - \frac{\partial J(u, m)}{\partial u} \left(\frac{\partial F(u, m)}{\partial u}\right)^{-1} \frac{\partial F(u, m)}{\partial m} + \frac{\partial J(u, m)}{\partial m}.
\end{equation}
Take the adjoint of the above equation:
\begin{equation}
\frac{d \widehat{J}(m)}{d m}^*  = - \frac{\partial F(u, m)}{\partial m}^* \left(\frac{\partial F(u, m)}{\partial u}\right)^{-*} \frac{\partial J(u, m)}{\partial u}^* + \frac{\partial J(u, m)}{\partial m}^*.
\end{equation}
Now define the new variable as
\begin{equation}
\lambda = \left(\frac{\partial F(u, m)}{\partial u}\right)^{-*} \frac{\partial J(u, m)}{\partial u}^*.
\end{equation}
This new variable is called the adjoint variable and it is the solution of the adjoint equation:
\begin{equation}\label{eq:adjoint1}
\frac{\partial F(u, m)}{\partial u}^{*} \lambda = \frac{\partial J(u, m)}{\partial u}^*.
\end{equation}
The adjoint equation is always linear.

\end{document}